\documentclass[10pt,twocolumn,letterpaper]{article}

\usepackage{cvpr}              %
\usepackage{csquotes}
\usepackage{mdframed}
\usepackage{tcolorbox}

\usepackage{float}

\definecolor{cvprblue}{rgb}{0.21,0.49,0.74}

\usepackage[pagebackref,breaklinks,colorlinks,allcolors=cvprblue]{hyperref}

\usepackage{array}
\usepackage{booktabs}
\usepackage{ragged2e}

\def\paperID{2549} %
\def\confName{CVPR}
\def\confYear{2025}

\title{
POSTA: A Go-to Framework for Customized Artistic Poster Generation
}

\author{
Haoyu Chen$^{1\dagger}$,\quad Xiaojie Xu$^{1\dagger}$, \quad Wenbo Li$^{2}$, \quad Jingjing Ren$^{1}$, \quad Tian Ye$^{1}$,\\ \vspace{1.2mm} 
Songhua Liu$^{3}$,\quad Ying-Cong Chen$^{1,4}$, \quad Lei Zhu$^{1,4*}$, \quad Xinchao Wang$^{3}$ \\ \vspace{-0.5mm}
\footnotesize $^{1}$The Hong Kong University of Science and Technology (Guangzhou)\quad
\footnotesize $^{2}$The Chinese University of Hong Kong\\ \vspace{-0.5mm}  
\footnotesize $^{3}$National University of Singapore \quad
\footnotesize $^{4}$The Hong Kong University of Science and Technology\\
{\tt\small Project page: \url{https://haoyuchen.com/POSTA}}\\
\footnotesize $^\dagger$Equal contribution \quad $^*$Corresponding author
}

\begin{document}

\twocolumn[{%
\renewcommand\twocolumn[1][]{#1}%
\maketitle
\vspace{-7mm}

\includegraphics[width=1\linewidth]{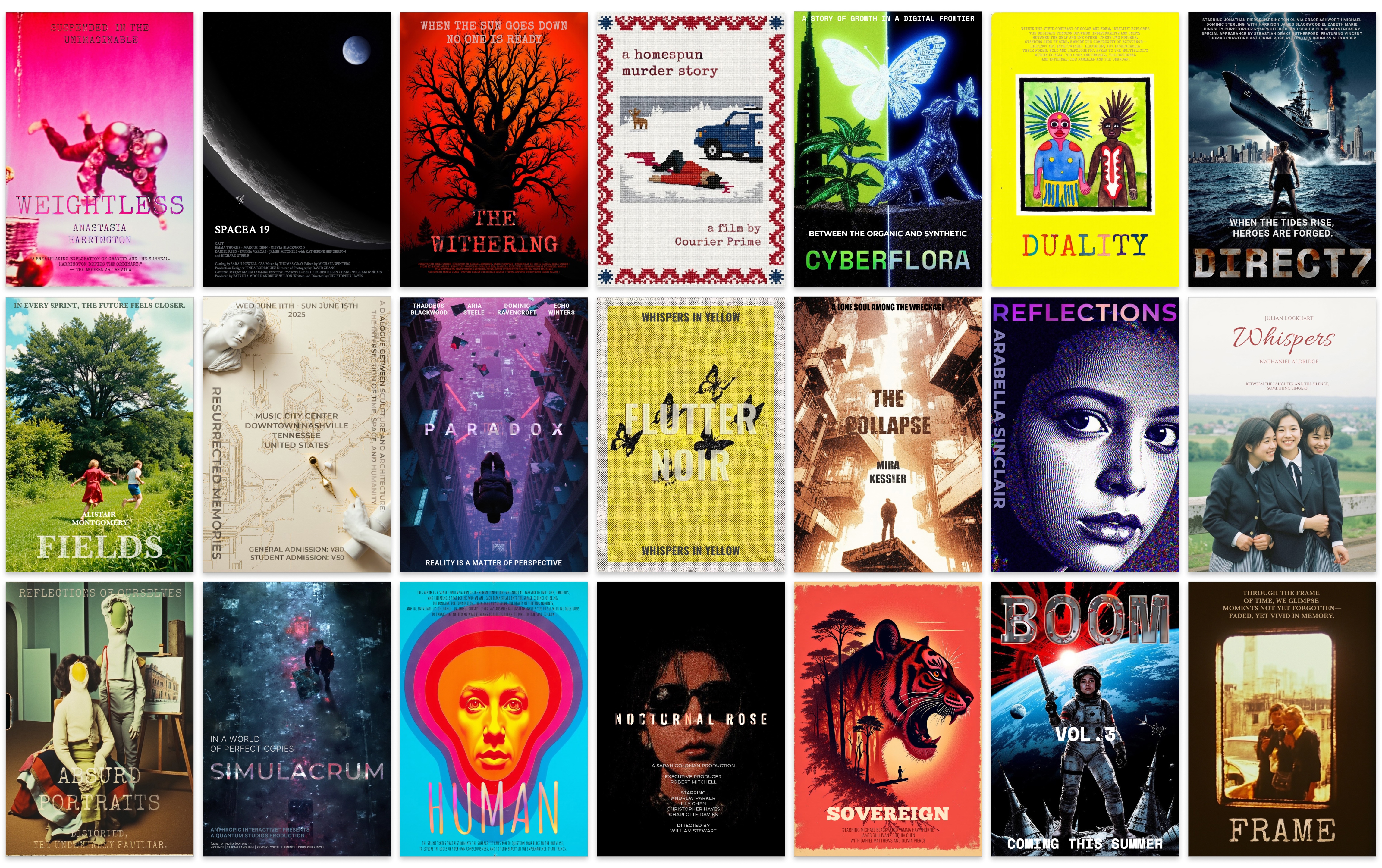}
\vspace{-7mm}
\captionof{figure}{Generated results using our \textit{POSTA} framework. The background, layout, and typographical designs are fully crafted from text inputs, showcasing the framework's capability to produce cohesive and visually engaging elements solely through textual guidance.}
\vspace{0.5em}
\label{fig:teaser}
}]

\begin{abstract}
\vspace{-9mm}

Poster design is a critical medium for visual communication. Prior work has explored automatic poster design using deep learning techniques, but these approaches lack text accuracy, user customization, and aesthetic appeal, limiting their applicability in artistic domains such as movies and exhibitions, where both clear content delivery and visual impact are essential.
To address these limitations, we present POSTA: a modular framework powered by diffusion models and multimodal large language models (MLLMs) for customized artistic poster generation. The framework consists of three modules. 
Background Diffusion creates a themed background based on user input. Design MLLM then generates layout and typography elements that align with and complement the background style. Finally, to enhance the poster's aesthetic appeal, ArtText Diffusion applies additional stylization to key text elements. The final result is a visually cohesive and appealing poster, with a fully modular process that allows for complete customization.
To train our models, we develop the \textit{PosterArt} dataset, comprising high-quality artistic posters annotated with layout, typography, and pixel-level stylized text segmentation.
Our comprehensive experimental analysis demonstrates \textit{POSTA}’s exceptional controllability and design diversity, outperforming existing models in both text accuracy and aesthetic quality.

\end{abstract}

\begin{figure*}[t]
  \centering
   \includegraphics[width=1\linewidth]{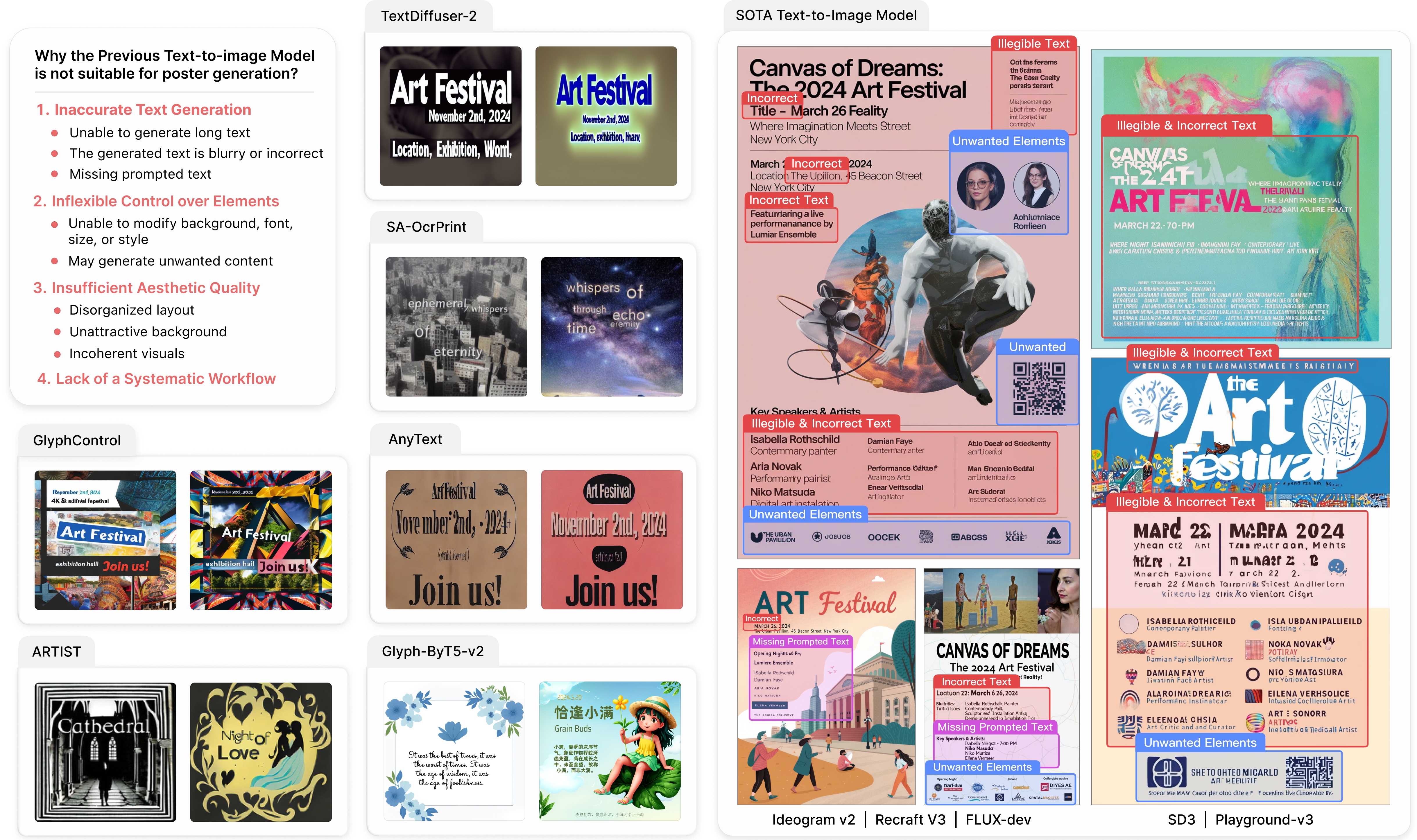}
   \vspace{-7mm}
   \caption{Our motivation stems from limitations of current methods for poster generation, which often struggle with issues like text inaccuracy, limited customization, and insufficient aesthetic quality.}
   \vspace{-5mm}
   \label{fig:onecol}
\end{figure*}

\vspace{-7mm}
\section{Introduction}
\label{sec:intro}
\vspace{-2mm}

Poster design plays a pivotal role in visual communication, especially in advertising, education, and the arts. Designers must thoughtfully integrate background, layout, color schemes, and typography to create compelling visuals. 
Generative models and transformers have already found numerous applications~\cite{li2023styo,qiu2023diffbfr,li2023dropkey,song2022transformer,luo2024diffusiontrack,li2024tokenpacker,yu2025inst3d,xiao2024asymmetric,xiao2024event,xiao2023cutmib,wu2025adaptive,wu2023image,wu2024prompt,chen2025restoreagent,chen2024low,chen2023snow,sun2024coser,ren2024ultrapixel,ren2025triplane,chen2023sparse,chen2024teaching,chen2023cplformer,10.1145/3655497.3655500,xiao-etal-2024-analyzing,10.1145/3696271.3696292}.
While many studies have explored automatic poster design using planning and generative models~\cite{glyph, textdiffuser, anytext, glyphcontrol, sd3, liu2024playground}, challenges in stability, customization, and flexibility prevent their widespread use in artistic domains such as concerts, movies, and exhibitions—areas that demand both clear content delivery and striking aesthetics. We propose that an effective poster design framework for these artistic fields must meet three key criteria: \textbf{(1) Text Accuracy}: Text must be precise and clear, as it is crucial for conveying the intended message. \textbf{(2) Customization}: Design elements should be fully editable and customizable to fit specific needs. \textbf{(3) Aesthetics}: The composition should be visually harmonious, ensuring that the design elements complement the background and overall aesthetic.

Current methods face significant challenges in meeting these criteria, as illustrated in \figurename~\ref{fig:onecol}, including: \textbf{(1) Inaccurate Text Generation:} Existing models struggle with generating accurate text, especially for long or complex content. They often produce misspellings or distorted characters, and even when bounding boxes are provided for text placement, the accuracy of longer text remains unreliable, limiting their practical use in real-world applications. \textbf{(2) Inflexible Control over Design Elements:} While many generative approaches can produce complete posters, their ``one-shot generation'' paradigm lacks the fine-grained control essential for professional design. Adjusting specific elements, such as background, text positioning, or font choices, is often cumbersome and inefficient. \textbf{(3) Insufficient Aesthetic Quality:} Existing methods tend to prioritize functionality over artistry, resulting in generated posters that lack the professional visual appeal necessary for commercial use. \textbf{(4) Lack of a Systematic Workflow:} Most current approaches focus on optimizing individual aspects of poster design, such as layout or typography, without addressing the full design process holistically. This fragmented approach makes it difficult to achieve a seamless, comprehensive workflow for poster creation.

To address these challenges, we propose POSTA, a modular framework for end-to-end intelligent poster creation. Combining generative models with Multi-Modal Large Language Models (MLLMs), POSTA offers a highly controllable and customizable solution. It overcomes key limitations through the integration of three components: \textbf{(1) Background Generation Module: } Utilizing a state-of-the-art diffusion model and several meticulously trained LoRA models~\cite{hu2021lora}, this module generates professional-grade backgrounds. Users also have the option to upload custom backgrounds, providing significant flexibility. \textbf{(2) Design Planning Module: } Leveraging MLLMs, this module intelligently plans the layout, text placement, and font attributes. The generated text is rendered based on these predictions, ensuring 100\% accuracy. Additionally, all text elements are fully editable, allowing for efficient customization. \textbf{(3) Artistic Text Stylization Module: } To enhance stylistic diversity and aesthetic appeal, we introduce a module that employs a mask-guided inpainting model~\cite{ju2024brushnet} to seamlessly integrate artistic text effects, ensuring visual harmony with the background.
The entire workflow aligns with the pipeline of professional designers.

To support design planning and artistic text stylization, we introduce the \textit{PosterArt} dataset, which is comprised of two parts: \textit{PosterArt-Design}, containing high-quality artistic posters with layouts, typographic elements, and backgrounds, and \textit{PosterArt-Text}, including pixel-wise segmentation, descriptions, and visual results of artistic text.

In summary, this work presents a systematic solution that can be seamlessly integrated into real-world design workflows, enabling designers to produce professional-quality outputs with ease. Our contributions are as follows. 
\begin{itemize}
\item We propose a systematic, modular workflow for highly controllable and reliable poster generation at a professional level, enabling users to easily adjust every aspect of the poster—including background, layout, and text—ensuring full customization.
\item We introduce effective text rendering and stylization modules that guarantee accurate text generation while enhancing the overall aesthetic appeal of the poster.
\item We demonstrate the versatility of our approach across various poster creation tasks, including text-guided poster generation, reference-based generation, and poster editing. Extensive experimental results validate the superiority of our method over existing methods.
\item We contribute the \textit{PosterArt} dataset, a high-quality, professionally curated resource comprising two subsets: \textit{PosterArt-Design}, which focuses on layout and typography, and \textit{PosterArt-Text}, which provides artistic text stylization and segmentation.
\end{itemize}

\vspace{-2mm}
\section{Related Work}
\label{sec:related_work}
\vspace{-1mm}

\subsection{Text-Aware Image Generation}
\vspace{-1mm}

Text-aware image generative models offer advantages for text-centric creative applications such as logo and simple poster design.
AnyText~\cite{anytext} introduced a diffusion pipeline with an auxiliary latent module and text embedding 
using OCR-based stroke encoding.
Similarly, GlyphControl~\cite{glyphcontrol} enhanced Stable Diffusion~\cite{podell2023sdxl,ldm} with glyph instructions, while Glyph-ByT5-v2~\cite{glyph} expanded multilingual text rendering through step-aware preference learning.
ARTIST\cite{zhang2024artist} used a disentangled architecture to capture text structure features, and TextDiffuser-2~\cite{textdiffuser} integrated large language models for layout modification through conversational interaction.
limiting user adjustments—crucial in professional design. Additionally, text rendering remains error-prone with complex typography. Our modular approach addresses this by ensuring text accuracy through typography prediction prior to artistic stylization.

\vspace{-1mm}
\subsection{Poster Layout Planning}
\vspace{-1mm}

Poster layout planning has advanced from simple templates to sophisticated learning-based systems.
PosterLayout introduced content-aware layouts with GAN-based design sequence formation for arranging elements. RADM~\cite{li2023relation} refined it through a relation-aware diffusion model that aligns visual, textual, and geometric relationships.
The integration of MLLMs~\cite{fu2024vita, zhang2024omg, wang2024qwen2, yao2024minicpm, ye2024mplug, tong2024cambrian} has advanced layout generation~\cite{yang2024mastering,hu2024ella,luo2024layoutllm}. PosterLLaVa~\cite{yang2024posterllava} introduced a MLLM framework, enabling flexible layout planning based on visual and textual constraints through language specifications.
These methods focus on basic layout prediction, overlooking key typographic elements like font, size, and orientation while \textit{POSTA} predicts these attributes jointly, ensuring a more cohesive and consistent design.

\begin{figure*}[t]
  \centering
   \includegraphics[width=1\linewidth]{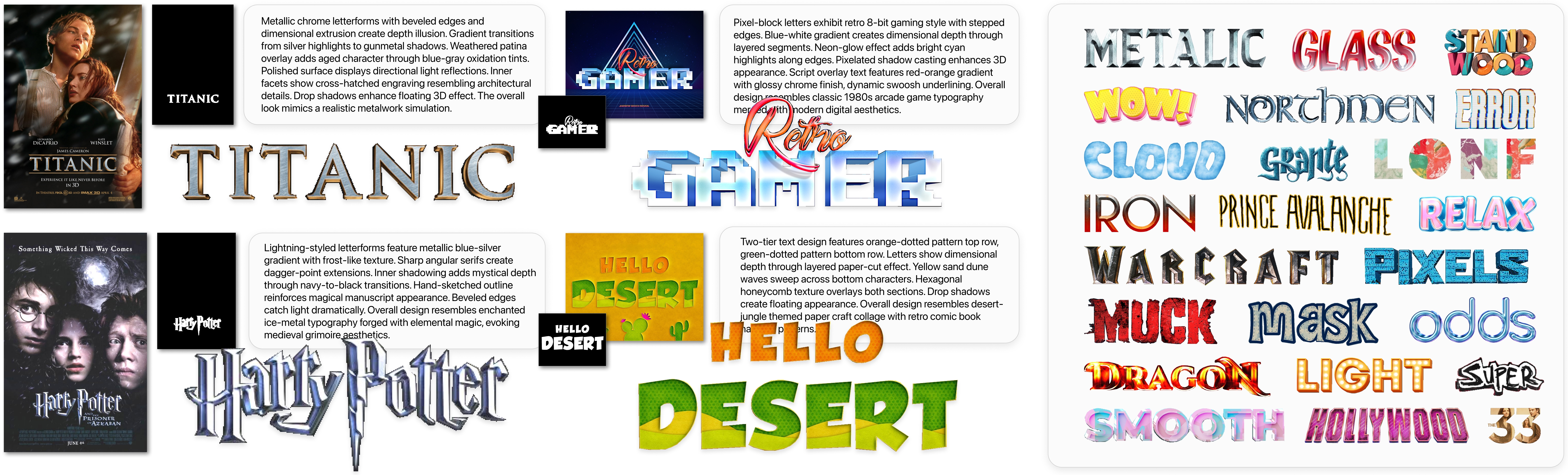}
   \vspace{-6mm}
   \caption{Overview of \textit{PosterArt-Text} dataset. It contains extensive segmentation and corresponding descriptions of texts with diverse artistic styles, primarily sourced from artistic posters such as those for movies, album covers, and similar media.}
   \vspace{-4mm}
   \label{fig:dataset_arttext}
\end{figure*}

\vspace{-1mm}
\subsection{Artistic Text Generation}
\vspace{-1mm}

Artistic text generation serves as a powerful tool for automating the creation of artistic visuals.
Approaches evolving from traditional stroke or patch-based methods to deep learning techniques like Generative Adversarial Networks~\cite{li2017universal,xu2021drb,song2021agilegan,karras2019style,park2019semantic,karras2020analyzing,zhang2017adversarial, tao2022df}.
Recently, large-scale models like Anything-to-Glyph~\cite{wang2023anything} and WordArt~\cite{he2023wordart} have enabled the generation of artistic text images. 
A detailed review is in~\cite{bai2024intelligent}. 
Although these methods create vivid text effects, they lack background context—for example, vintage posters often feature worn text to convey age. Our text stylization is trained with context, ensuring text effects harmonize with the style and semantics of the background.

\vspace{-1mm}
\subsection{Modular Graphic Design}
\vspace{-1mm}
The complete poster design process encompasses background generation, layout planning, and text rendering. However, existing approaches typically address only one of these components. To our knowledge, COLE\cite{jia2024colehierarchicalgenerationframework} is the only framework to propose a comprehensive workflow for this task. 
The COLE system integrates multiple fine-tuned LMMs, and Diffusion Models, each tailored for design-aware layer-wise captioning, layout planning, reasoning, and image and text generation. However, its multi-step process is complex and requires extensive intermediate checks, making it time-consuming. OpenCOLE\cite{inoue2024opencole} simplifies some components but is trained on datasets outside the artistic domain, limiting its ability to generate aesthetically pleasing results.
\textit{POSTA} streamlines layout and typographic elements, leveraging a specialized artistic text stylization module to create harmonious, visually engaging text. Powered by curated data tailored for artistic posters, our results significantly surpass those of concurrent approaches.

\vspace{-2mm}
\section{PosterArt Dataset}
\vspace{-1mm}
\label{sec:dataset}

To support the development of our framework and enhance its ability to generate high-quality, professional poster designs, we introduce the \textit{PosterArt} dataset. This dataset is designed specifically for poster creation tasks and is divided into two subsets: \textit{PosterArt-Design}, which focuses on layout and typography, and \textit{PosterArt-Text}, which is dedicated to artistic text stylization.

\vspace{-5mm}
\paragraph{PosterArt-Design}

\begin{figure}[t]
  \centering
   \includegraphics[width=1\linewidth]{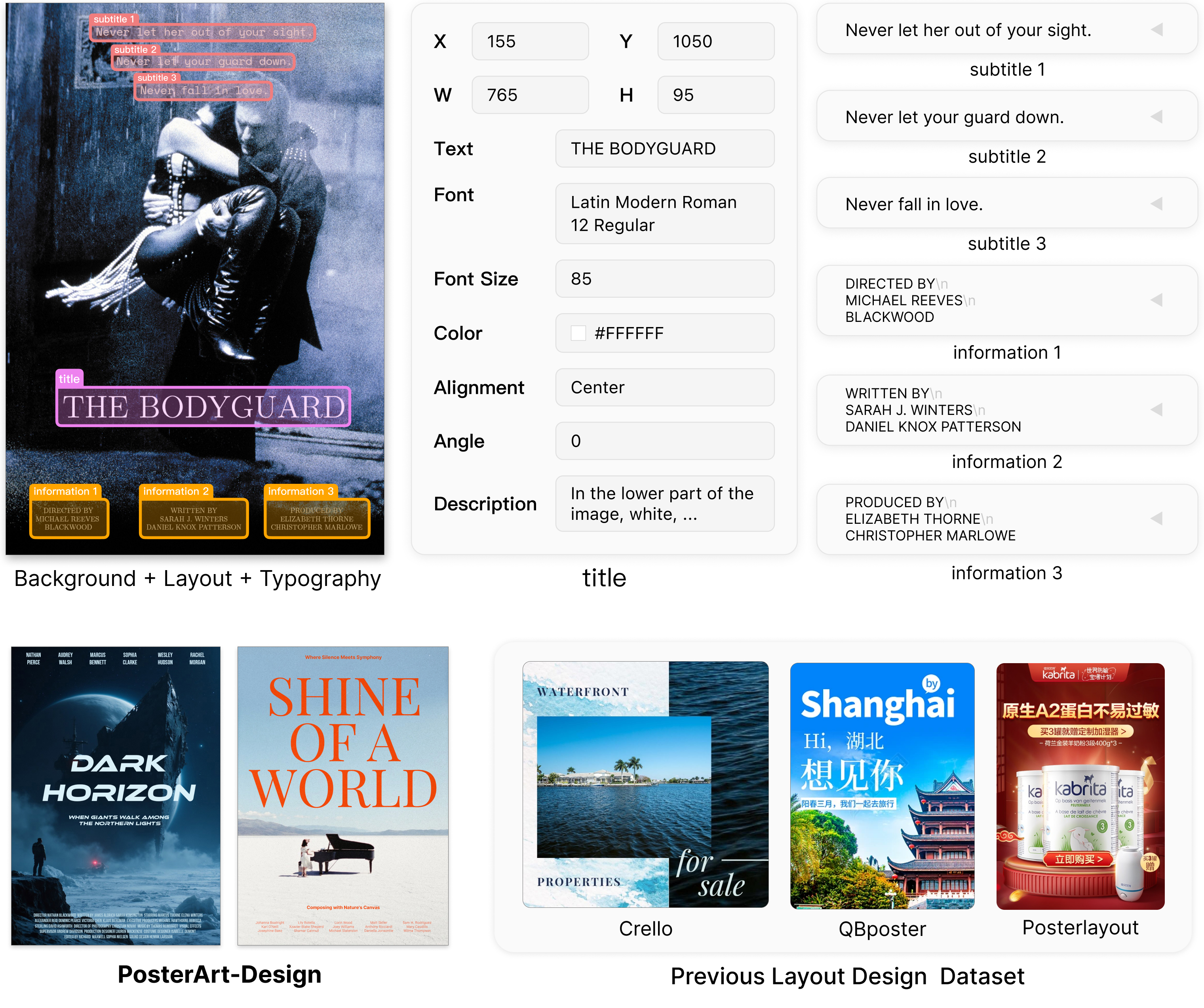}
   \vspace{-7mm}
   \caption{A sample of \textit{PosterArt-Design} (top). \textit{PosterArt-Design} vs. previous layout datasets (bottom). Our dataset is crafted by expert designers who carefully incorporate elements into backgrounds, including deliberate layout and typography information. }
   \vspace{-5mm}
   \label{fig:dataset}
\end{figure}

To facilitate the generation of visually compelling, professionally designed poster layouts and typography, we developed \textit{PosterArt-Design}.
Specifically, \textit{PosterArt-Design} consists of over 2,000 poster backgrounds collected from the Internet, with additional aesthetically pleasing layout and typography information crafted by professional designers. For each text element, we annotated key attributes—including coordinates, font type, font size, color, alignment, and rotation angle, as shown in \figurename~\ref{fig:dataset}.
These annotations provide essential design information for fine-tuning MLLMs, enabling guided text placement and style generation. A brief user description accompanies each element to further inform the model.

In contrast to previous layout datasets, which comprise commercial ads or randomly sourced social media posters (\figurename~\ref{fig:dataset}, bottom right) that often feature disorganized and aesthetically inferior designs lacking essential typographic elements like font type, size, and color, our dataset is curated to include only high-quality artistic posters crafted by professional designers. Each poster in our dataset exemplifies a high aesthetic standard and includes key typographic elements critical for effective design.

\begin{figure*}[t]
  \centering
   \includegraphics[width=1\linewidth]{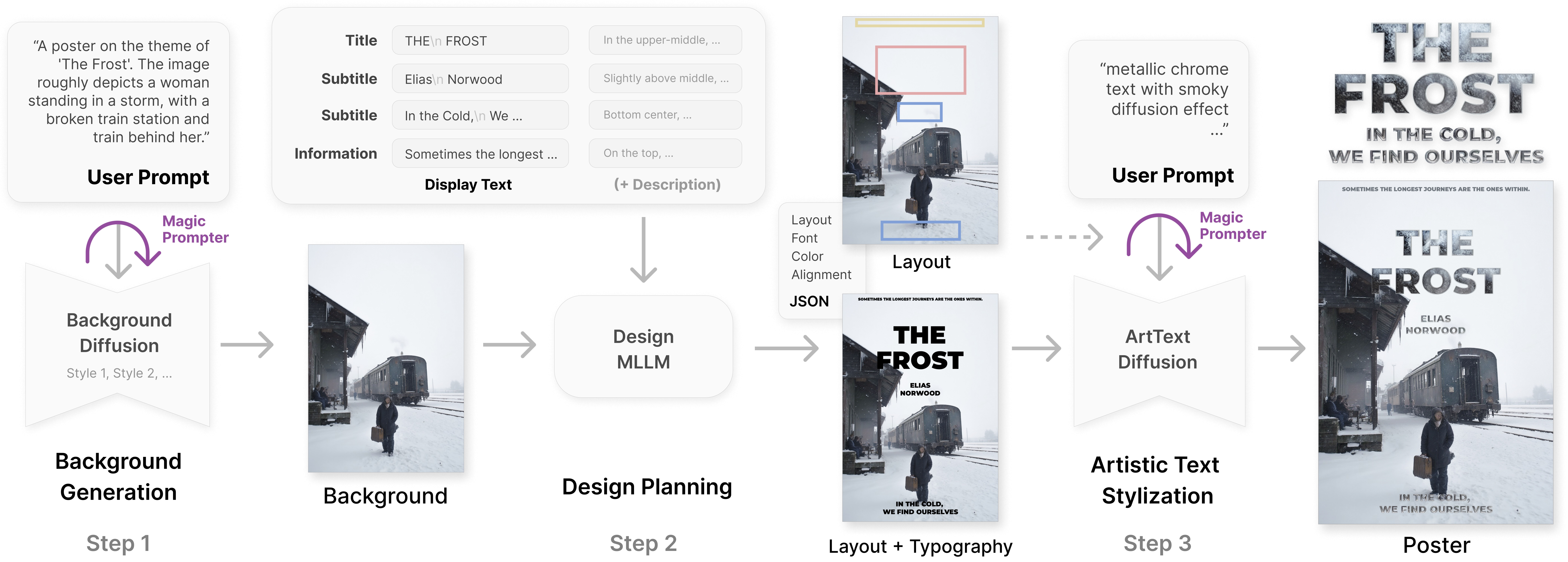}

   \vspace{-2mm}
   \caption{Our \textit{POSTA} pipeline consists of three steps: background generation, design planning, and artistic text stylization. Background Diffusion and ArtText Diffusion are employed to generate backgrounds and text with artistic effects, while the Design MLLM predicts layout and typography information. The GPT-4V-powered Magic Prompter is used to refine prompts based on user descriptions or background images, optimizing input for the diffusion models.}
   \vspace{-5mm}
   \label{fig:method}
\end{figure*}

\vspace{-3mm}
\paragraph{PosterArt-Text}

\textit{PosterArt-Text} focuses on the artistic text regions in poster titles, as illustrated in \figurename~\ref{fig:dataset_arttext}. This dataset aims to empower models to generate text that is both visually appealing and seamlessly integrated with the poster's background and design elements, ensuring a harmonious and cohesive appearance.
We created this dataset by collecting a large number of visually compelling posters, generating descriptive captions via GPT-4V\cite{gpt4v}, and manually segmenting title regions. To enhance diversity, we used FLUX\cite{flux} to generate text with artistic effects, such as "baroque font made of coral with pearly shine" In total, the dataset comprises about 2,500 detailed text segmentations and descriptions harmonized with their backgrounds.

A notable contribution of \textit{PosterArt-Text} is its dual functionality for both artistic text generation and segmentation tasks. 
Even advanced segmentation models often struggle to accurately isolate complex title fonts in posters. By providing high-quality, human-annotated segmentation masks, this dataset offers a valuable resource for improving text segmentation accuracy, especially when the text appears naturally integrated with complex or intricate backgrounds.

\vspace{-1mm}
\section{Method}
\vspace{-1mm}
\label{sec:method}

Our \textit{POSTA} pipeline, as illustrated in \figurename~\ref{fig:method}, consists of three key components: Background Diffusion, Design MLLM, and ArtText Diffusion. Users provide descriptive inputs at each stage, and the framework generates the background image, layout and typography details, and the final poster featuring artistic text elements.
The entire process is fully controllable, allowing for customization at each stage.

\begin{figure*}[t]
  \centering
   \includegraphics[width=1\linewidth]{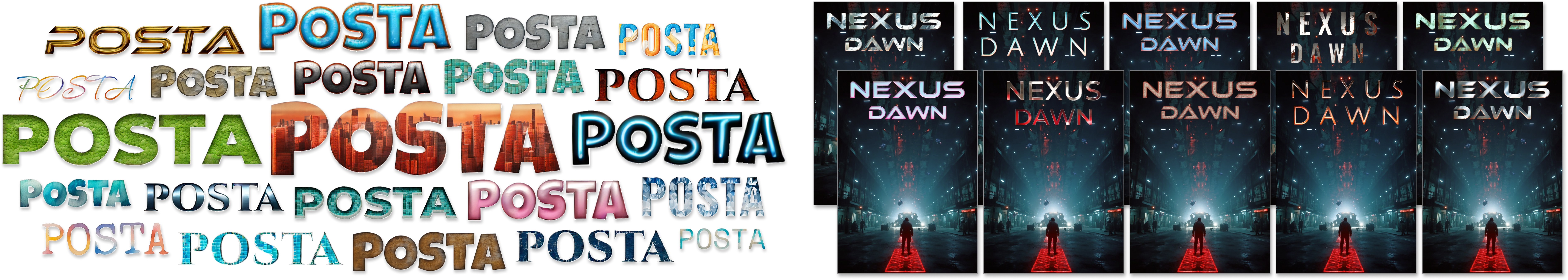}
   \vspace{-6mm}
   \caption{The artistic texts generated by our framework are highly diverse, with each style crafted to suit a specific theme.}
   \vspace{-5mm}
   \label{fig:font}
\end{figure*}

\vspace{-1mm}
\subsection{Background Generation}
\vspace{-1mm}

A high-quality art poster begins with a well-designed background. In this stage, Background Diffusion, built upon FLUX, is capable of generating a diverse range of premium background images. To improve accuracy across various artistic styles, we collect poster samples categorized into distinct styles such as minimalism, retro, and modern art. These samples are used to train multiple specialized Low-Rank Adaptation (LoRA) blocks, each responsible for controlling specific stylistic elements. To ensure optimal outputs, we integrate an MLLM-powered Magic Prompter to refine user input prompts by adding details.

\vspace{-1mm}
\subsection{Design Planning}
\vspace{-1mm}

To fundamentally address the issue of imprecise text generation while meeting the dual requirements of editability and aesthetic quality, we develop a novel poster design planning method that seamlessly integrates layout, typography generation, and text rendering.
Effective text layout prediction requires a deep understanding of the user's design intent and the semantic relationship between text and images. To achieve this, we leverage an MLLM, using LLaVA\cite{liu2024visual} as the backbone of our design model. Through visual instruction tuning, the model plans the layout and typography for each text element based on user input. Text elements are categorized into three types—title, subtitle, and information—reflecting their prominence in the design. Users can specify text along with optional attributes such as position, alignment, or layout style. Additionally, users may provide high-level layout descriptions to guide the relative positioning of elements. The model outputs key attributes, including position, size, font, color, alignment, and rotation angle, as shown in \figurename~\ref{fig:dataset}.

For text rendering, to completely resolve the issue of imprecise text generation in generative models, we directly render text based on the font details provided by the model, rather than generating text images. This approach aligns with professional design workflows. By storing text information in an editable vector format, our system enables users to fine-tune design elements such as size, position, font, and alignment after generation.

\vspace{-2mm}
\subsection{Artistic Text Stylization}

To further enhance the visual impact of the text—particularly the poster title—and ensure seamless integration with the background while preserving text accuracy, we employ a mask-guided inpainting strategy. The ArtText Diffusion model is based on BrushNet~\cite{ju2024brushnet}, a ControlNet-like~\cite{zhang2023adding} architecture designed for the inpainting task. Trained on our high-quality \textit{PosterArt-Text} dataset, this model is capable of understanding the textual descriptions of font styles and generating text with artistic features such as 3D effects, metallic textures, color gradients, and outlines. Guided by text prompts generated according to user input or the background style with Magic Prompter, the model performs local generation to ensure that the text naturally blends with the background. The blending process could be written as:
{\abovedisplayskip=3pt
\belowdisplayskip=3pt
\begin{equation}
I_{\text{blended}} = M \odot I_1 + (1 - M) \odot I_2
\end{equation}}
where $M$, $I_1$, and $I_2$ denote the mask, generated image, and original image, respectively. A Gaussian kernel is applied to the mask to achieve smooth boundaries. The entire process achieves stylized artistic text effects that harmonize with the overall design while ensuring text accuracy.

\vspace{-1mm}
\section{Application}

\vspace{-1mm}
\subsection{Text-Based Poster Generation}
\vspace{-1mm}

\figurename~\ref{fig:teaser} demonstrates the comprehensive capabilities of our \textit{POSTA} framework in generating diverse poster designs. Our system exhibits remarkable versatility in producing various types of visual content, including book covers, movie posters, promotional materials, and event announcements, all generated exclusively from textual descriptions.

\vspace{-1mm}
\subsection{Artistic Text Generation}
\vspace{-2mm}

One critical components of our \textit{POSTA} framework is the ArtText Diffusion module, which enhances aesthetic appeal of the generated posters. This feature allows us to craft unique and visually engaging fonts that go beyond traditional typography, enhancing the overall expressiveness of the designs. As shown in \figurename~\ref{fig:font}, our method is capable of producing a wide variety of artistic fonts, each designed to suit different stylistic themes and visual contexts.
\vspace{-3mm}
\begin{figure}[H]
  \centering
   \includegraphics[width=1\linewidth]{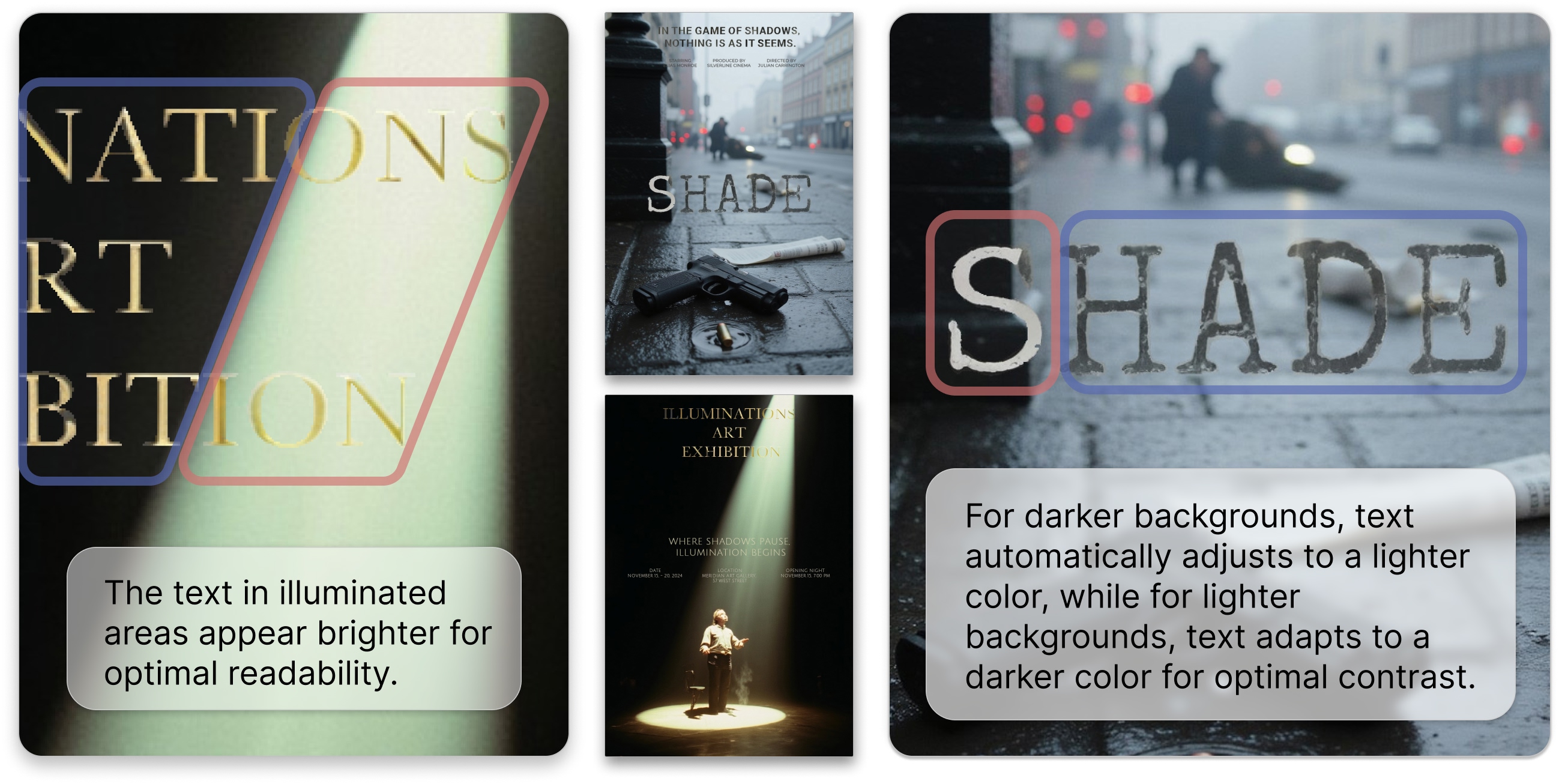}
   \vspace{-6mm}
   \caption{In various background regions, text displays distinct illumination and color effects.}
   \vspace{-5mm}
   \label{fig:text_background}
\end{figure}

\begin{figure}[t]
  \centering
   \includegraphics[width=1\linewidth]{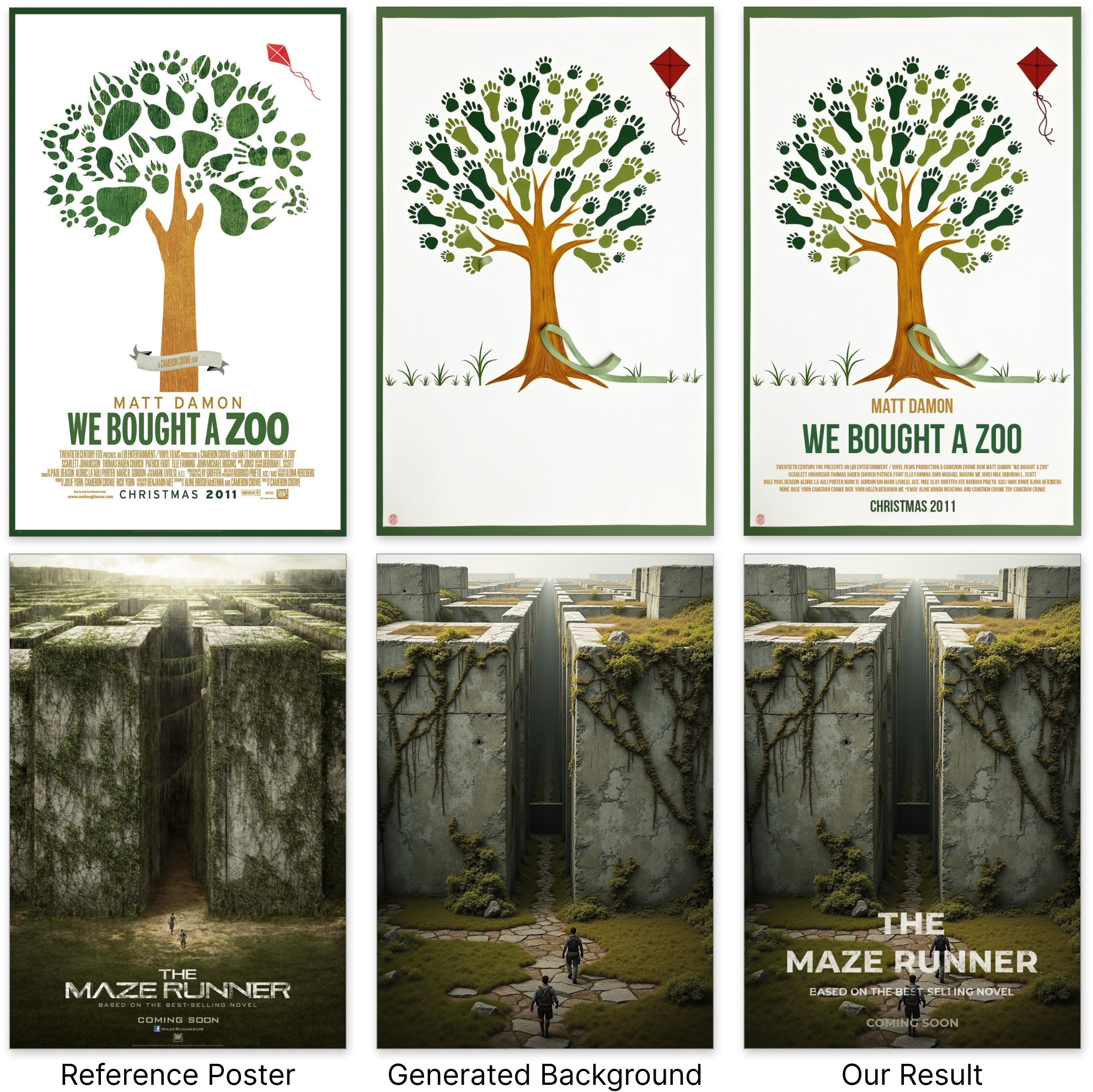}

   \vspace{-2mm}
   \caption{Reference-Based Poster Generation.}
   \vspace{-7mm}
   \label{fig:reference}
\end{figure}

Another key feature of \textit{POSTA} is its ability to seamlessly integrate fonts into the background, ensuring that the fonts do not appear disconnected. This capability sets our framework apart from previous models, which often struggle to harmonize text with the surrounding. As shown in \figurename~\ref{fig:text_background}, our system dynamically adjusts the font's color, brightness, and contrast to blend naturally with the background, significantly enhancing readability and visual coherence.%

\begin{figure*}[t]
  \centering
   \includegraphics[width=1\linewidth]{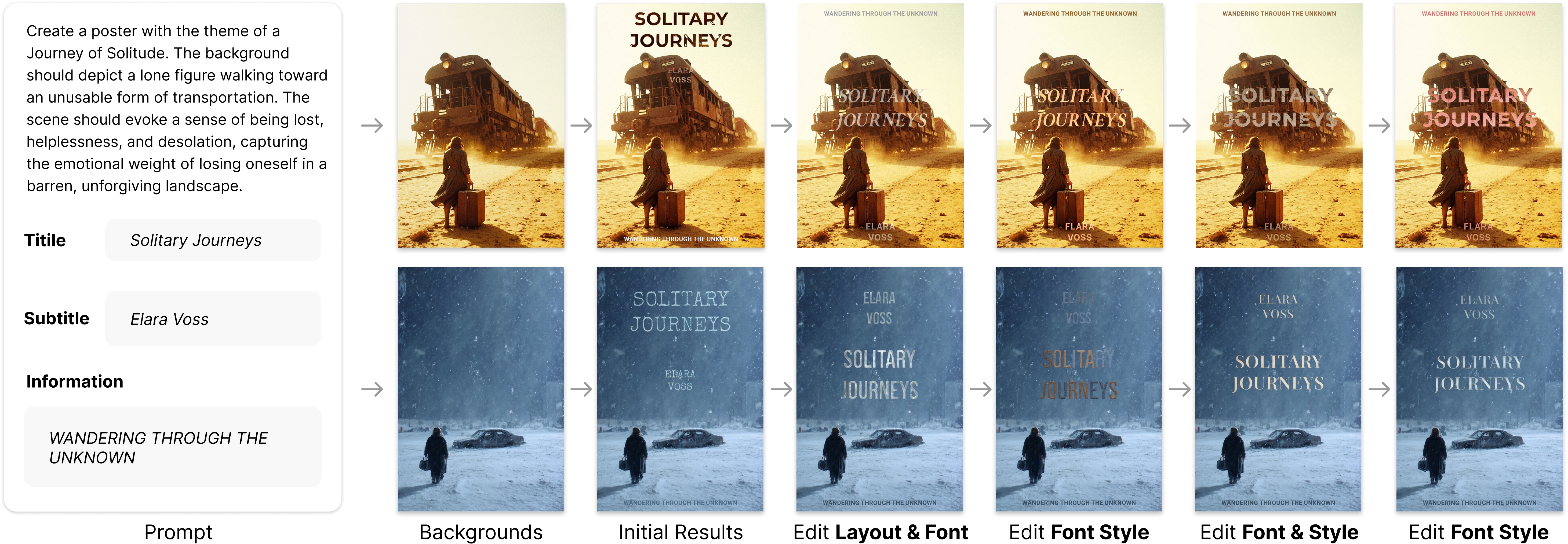}
   \vspace{-6mm}
   \caption{Our framework provides full editability over all design elements-layouts, fonts, styles, and backgrounds.}
   \vspace{-5mm}
   \label{fig:control}
\end{figure*}

\vspace{-2mm}
\subsection{Reference-Based Poster Generation}
\vspace{-2mm}

Our framework also enables Reference-Based Poster Generation, as demonstrated in \figurename~\ref{fig:reference}. The process starts with image-to-image transfer to remove text while preserving the background.  GPT-4V then extracts a detailed description of the reference, guiding the generation of a new layout and typography. Finally, the title is stylized to match the background, with the prompt generated by GPT-4V. This approach creates visually coherent posters that retain the essence of the reference while allowing full customization.

\vspace{-2mm}
\subsection{Demonstration of Editability}
\vspace{-2mm}

A key feature of our framework is its control and editability over design elements, allowing users to modify fonts, layouts, backgrounds, and styles to meet specific needs. As shown in \figurename~\ref{fig:control}, the system generates multiple poster variations from the same prompt, each with unique backgrounds and typography layouts. Users can make precise edits, such as adjusting layouts, changing fonts, or refining font effects (e.g., color, weight, texture). This flexibility ensures seamless text-background alignment, enabling iterative improvements toward professional-quality designs.

\vspace{-1mm}
\section{Experiment}

\subsection{Implementation Details}
\vspace{-1mm}

For Background Diffusion, we train FLUX LoRA\cite{hu2021lora} at a resolution of 1024, using approximately 50 images per style, with an embedding dimension of 64.
For Design MLLM, we use the CLIP pre-trained Vision Transformer (ViT-L/14)~\cite{clip} as the image encoder to convert input images into visual tokens, and Llama3 7B~\cite{llama} as the language model. Training was conducted on 2 A800 GPUs.
For ArtText Diffusion, we train based on SDXL's BrushNet~\cite{podell2023sdxl, ju2024brushnet} at a resolution of 1216. Further implementation details are provided in the supplementary materials.

\vspace{-1mm}
\subsection{Qualitative Results and Comparisons}
\begin{figure}[H]
  \centering
     \vspace{-2mm}
   \includegraphics[width=1\linewidth]{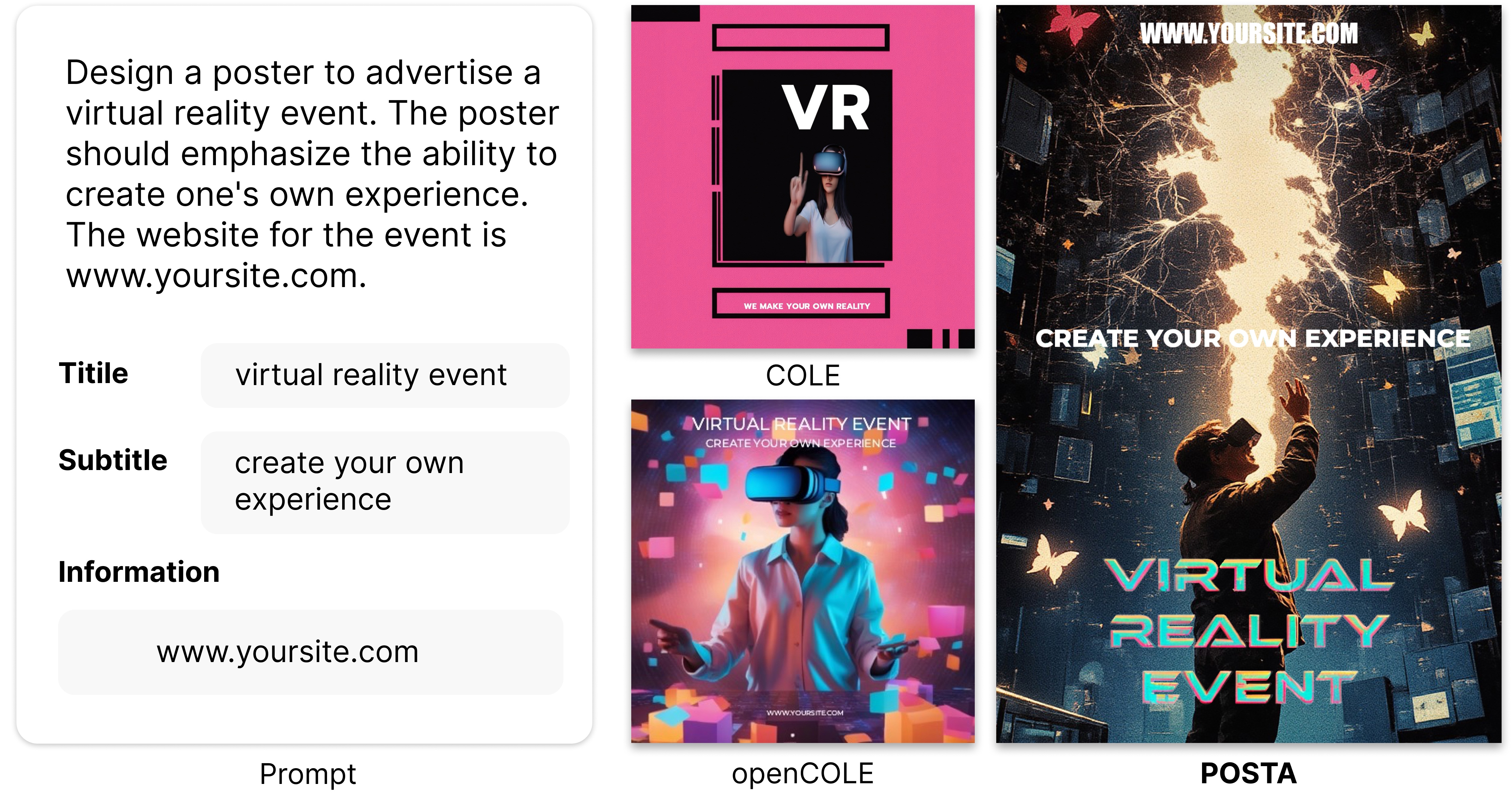}
   \vspace{-6mm}
   \caption{Comparison with COLE and OpenCOLE.}
   \vspace{-3mm}
   \label{fig:style}
\end{figure}

\begin{figure*}[t]
  \centering
   \includegraphics[width=1\linewidth]{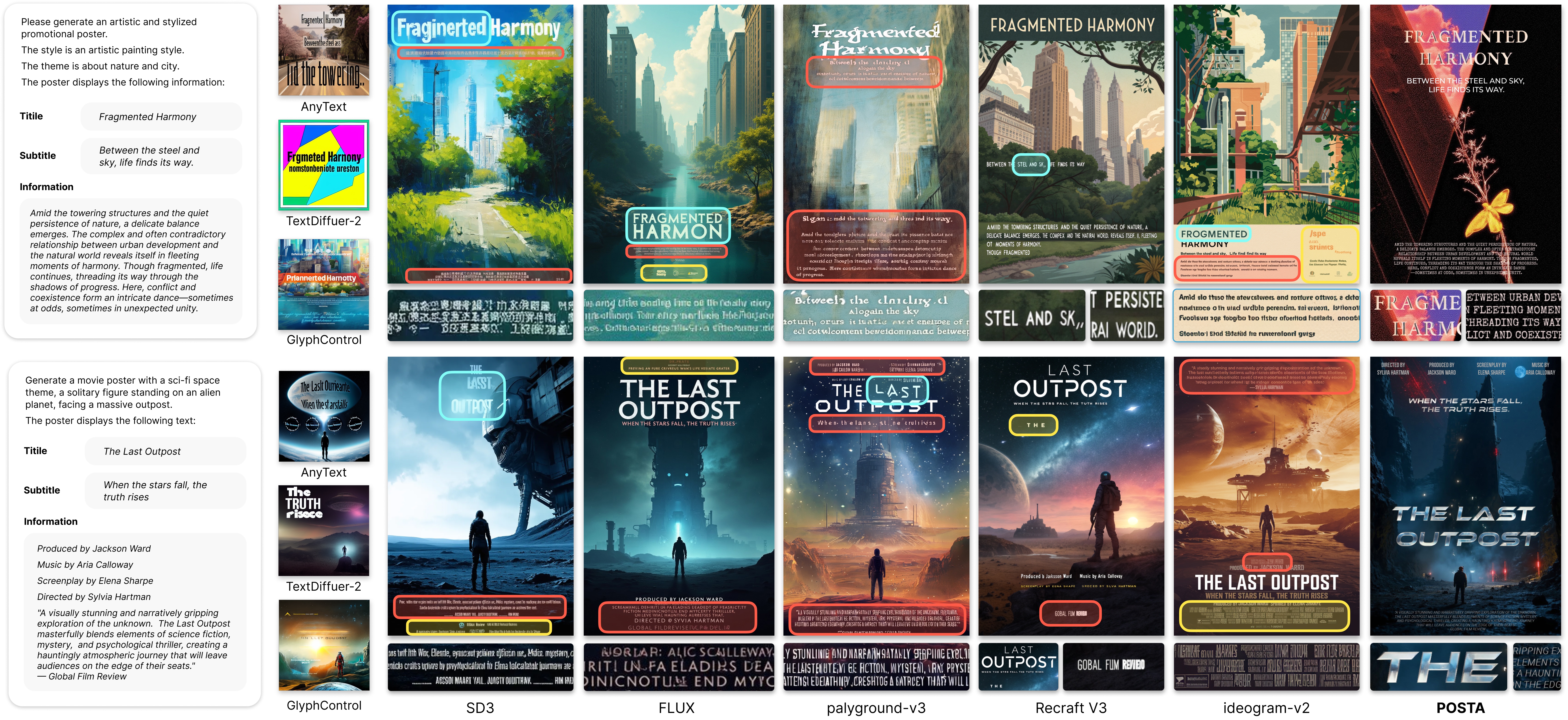}
   \vspace{-6mm}
   \caption{Comparisons with state-of-the-art models, encompassing popular open-source and proprietary commercial models.
\textcolor[HTML]{83FDF7}{Blue box}: word generation errors.
\textcolor[HTML]{FF5B48}{Red box}: unclear or unrecognizable text.
\textcolor[HTML]{FFE748}{Yellow box}: generated content that is irrelevant to the prompt.
}
   \vspace{-3mm}
   \label{fig:style2}
\end{figure*}

\vspace{-1mm}
To evaluate the effectiveness of our \textit{POSTA} framework, we conducted a comprehensive comparison with state-of-the-art models, including both popular end-to-end open-source and proprietary commercial models (as shown in \figurename~\ref{fig:style2}), along with research works on systematic poster design, namely COLE\cite{jia2024colehierarchicalgenerationframework} and OpenCOLE\cite{inoue2024opencole} (as shown in \figurename~\ref{fig:style}). As illustrated, \textit{POSTA} significantly outperforms the others on accurate text rendering, user description understanding and artistic stylization, particularly in handling long text sequences.

A major flaw in all other models is their inability to accurately generate even moderately long text. In most cases, the text output is not only incomplete but degenerates into garbled characters. Models such as AnyText~\cite{anytext}, TextDiffuser-2~\cite{textdiffuser}, and Flux frequently fail to render titles correctly, resulting in unreadable text. Moreover, these models often produce unwanted content, such as irrelevant symbols or extraneous marks that the user did not specify. This behavior leads to cluttered and unprofessional results, making them unsuitable for high-quality design tasks. In contrast, \textit{POSTA} maintains complete control over elements, ensuring that only the desired content appears, and that it is rendered with precision and clarity, even for longer text.

In addition to superior text accuracy, \textit{POSTA} also excels in artistic font quality. While other models struggle to generate visually appealing and well-integrated fonts, \textit{POSTA} produces fonts that are not only aesthetically pleasing but also harmonize with the overall design. This ensures that the text does not appear disjointed or out of place, as is often seen with other methods.

\begin{figure}[t]
  \centering
   \includegraphics[width=1\linewidth]{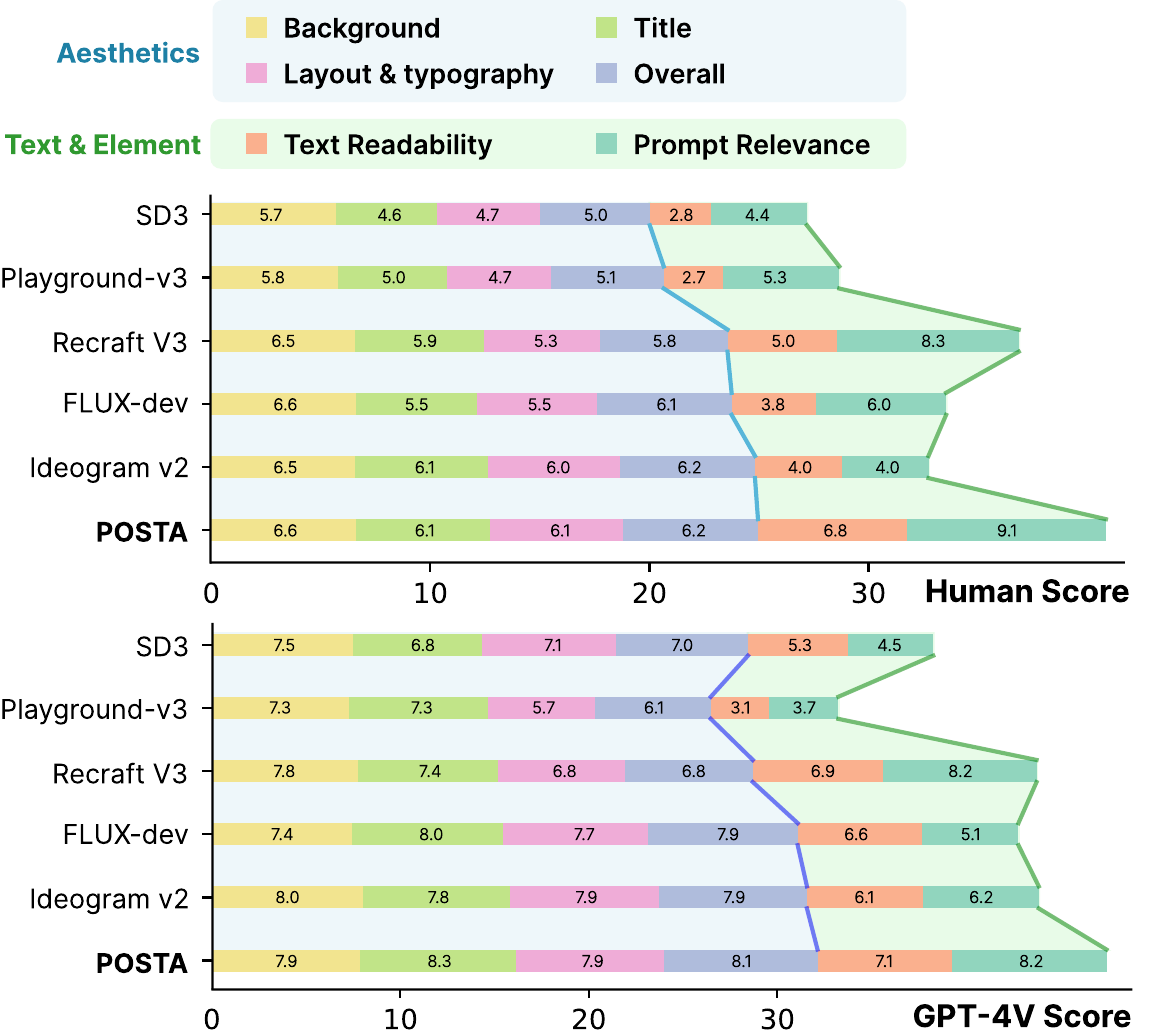}
   \vspace{-6mm}
   \caption{Human and GPT-4V evaluation. }
   \vspace{-6mm}
   \label{fig:userstudy}
\end{figure}

\vspace{-1mm}
\subsection{Quantitative Results and Comparisons}
\vspace{-1mm}
We conduct quantitative analysis comparing our \textit{POSTA} framework with state-of-the-art models, including 20 poster samples generated with the same prompts using advanced general-purpose models (SD3~\cite{sd3}, Playground-v3~\cite{liu2024playground}, FLUX-dev) and graphic design-specific models (Recraft V3\cite{recraft}, Ideogram v2~\cite{ideo}). Sixty users with experience in AI tools and an arts background, along with GPT-4V, assess key aspects such as visual appeal, text readability, and prompt-image relevance, rating from 1 to 10.

As shown in \figurename~\ref{fig:userstudy}, general-purpose models performed moderately, with FLUX-dev leading among them. Graphic design-specific models, Recraft and Ideogram, outperform open-source models in aesthetics but have notable issues. Recraft struggles with longer text sequences, often omitting specified text, while Ideogram introduces irrelevant elements that compromise design quality. In contrast, \textit{POSTA} excels across all criteria, delivering superior aesthetics, text readability, and prompt relevance, making it the most reliable option for professional poster generation.

\begin{figure}[t]
  \centering
   \includegraphics[width=1\linewidth]{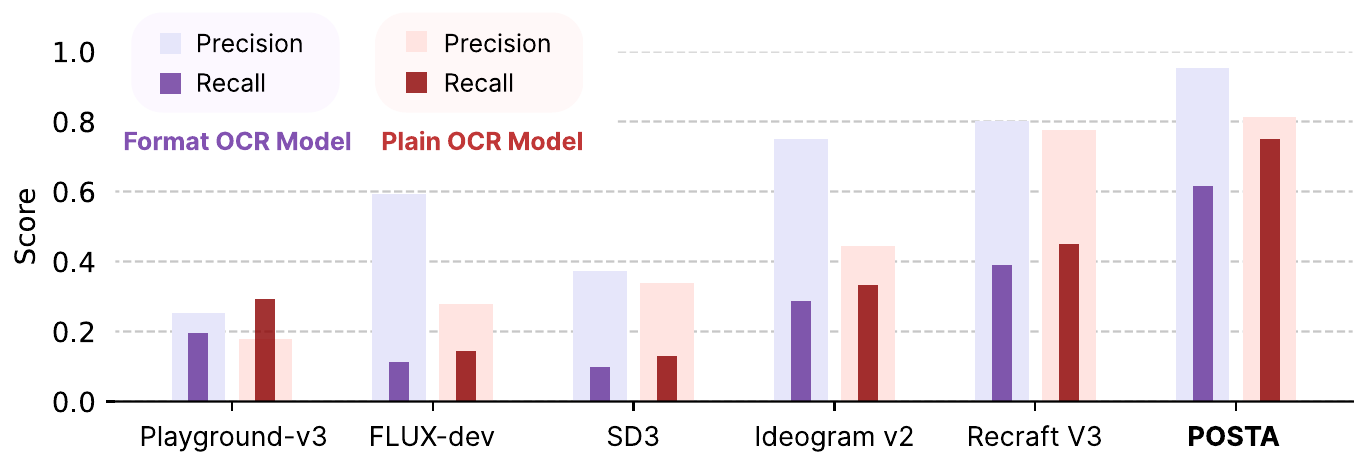}
   \vspace{-6mm}
   \caption{OCR-based comparisons, where "Format / Plain" indicates whether models are designed for electronic print or handwritten characters.}
   \vspace{-6mm}
   \label{fig:ocr}
\end{figure}

For text accuracy, we further employ GOT-OCR\cite{wei2024general} to detect text on generated posters, calculating the average word-level precision and recall between detected and user-specified text. As shown in \figurename~\ref{fig:ocr}, \textit{POSTA} achieves the highest level of text consistency with user input, since we generate absolutely accurate text.

\vspace{-3mm}
\section{Conclusion}
\vspace{-2mm}

In this paper, we present \textit{POSTA}, a framework designed to harness the capabilities of vision diffusion models and multimodal language models for customized artistic poster generation. Specifically, we employ two diffusion models—one for background generation and another for artistic text stylization—along with a multimodal language model for layout and typography planning. To train our models, we introduce \textit{PosterArt} dataset, comprising high-quality artistic posters annotated with layout, typography, and pixel-level stylized text segmentation. Experimental results confirm that \textit{POSTA} effectively produces highly diverse, visually compelling posters with high text accuracy.

As for limitations, our Design MLLM is currently limited to generating relatively simple layout designs and offers a restricted selection of font types, which constrains design diversity and limits the performance of the ArtText Diffusion. This limitation stems from the high standards in artistic data production. In the future,  expanded datasets and more advanced model architectures will enhance the scalability.

{
    \small
    \bibliographystyle{ieeenat_fullname}
    \bibliography{main}
}

\end{document}